\documentclass[prl,superscriptaddress,notitlepage,twocolumn,nofootinbib,a4paper]{revtex4-1}

\usepackage{amsmath}    
\usepackage{graphicx}   
\usepackage{verbatim}   
\usepackage{color}      
\usepackage{hyperref}   
\usepackage{amsfonts}
\usepackage[english]{babel}
\usepackage{ucs}
\usepackage[utf8x]{inputenc}
\usepackage{subfigure}
\usepackage[autostyle]{csquotes}
\usepackage{float}

\newcommand{\lra}[1]{\langle{}#1\rangle}

\begin{document}

\title{Collective biological computation of metabolic economy}

\author{D.~Korošak}
\thanks{Corresponding author:\\dean.korosak@um.si}
\affiliation{University of Maribor, Faculty of Medicine, Institute for Physiology, Maribor, Slovenia}
\affiliation{University of Maribor, Faculty of Civil Engineering, Transportation Engineering and Architecture, Maribor, Slovenia}

\author{A.~Stožer}
\affiliation{University of Maribor, Faculty of Medicine, Institute for Physiology, Maribor, Slovenia}

\author{M.~Slak Rupnik}
\thanks{Corresponding author:\\marjan.slakrupnik@meduniwien.ac.at}
\affiliation{Medical University of Vienna, Center for physiology and pharmacology, Vienna, Austria}
\affiliation{University of Maribor, Faculty of Medicine, Institute for Physiology, Maribor, Slovenia}
\affiliation{Alma Mater Europaea - European Center Maribor, Maribor, Slovenia}


\date{\today}

\begin{abstract}
Presented with sensory challenges, living cells employ extensive noisy, fluctuating signalling and communication among 
themselves to compute a physiologically proper response which often results in symmetry breaking. We propose, based on the results of a coupled stochastics oscillators model that biological computation mechanism undertaken by insulin secreting beta-cells consists of a combination of dual intracellular Ca$^{2+}$ release processes to ensure multilayered exploration contributing to enhanced robustness and sensitivity. The computational output is what is macroscopically observed as disorder-order phase transition in collective beta-cell response to nutrient concentration increase. Based on the analogies from previoulsy described examples of biological computation, we argue that the initial response may be followed by an adaptive phase to expand the sensory spectrum and consolidate memory.      
\end{abstract}

\maketitle 


\section{Introduction}
The concept that living cells compute is almost trivial, yet the very nature of biological computation itself is much harder to precisely define, model and describe. This is even more difficult when we search for these answers in analogies between computation preformed by a cell collective and by Turing machines or man-made digital computational architectures~\cite{mitchell2009complexity}. Biological systems operate in information rich, fluctuating and noisy environment from which they must extract clues in order to successfully respond and adapt to. Physiological processes that underlie such biological information processing~\cite{tkacik2016information} and decision making fundamentally depend on symmetry breaking~\cite{hopfield1994physics, li2010symmetry}, therefore it is posited that it would be beneficial if biological systems would reside near a critical point~\cite{mora2011biological} of a phase transition that would provide optimal conditions for biological computations~\cite{langton1990computation}.

Here we consider homeostatic regulation of nutrients and other metabolic intermediates in higher animals as one of many biological computational problems that organisms continously solve to ensure their survival and reproduction. The main task of this computation is to supply the metabolic code in the form of hormones like insulin or glucagon to economize with metabolites. The metabolites can be either invested into cell housekeeping to support integrity of the cell structure and function, they can be stored into celullar depots for the future use, or they are maintained in the circulating blood as a metabolic liquidity to pay for current or acutely emerging energetic requirements. This metabolic liquidity is maintained even during long periods of complete starvation~\cite{CahillNEJM1970}. On the other hand, a deleterious hormonal perturbation that leads to an exaggerated transport of glucose into the cells easily pushes an organism into hypoglycemia with glucose levels below 3.9 mmol/l. The resulting metabolic liquidity issues limit the function of neurons and blood cells, which by number represent a vast majority (90\%) of all cells in a human body, and lead to a major acute stress response with long-term consequences~\cite{HAAS2022102983}. The metabolic economy is therefore a key physiological process, driven by small pancreatic endocrine cell collectives, mostly insulin-secreting beta-cells. These cells are distributed all over the pancreas gland. They sense, communicate, compute and respond with a precise hormone release to metabolic changes using a combination of intra- and intercellular, as well as interislet communication mechanisms.

It is an open question how the abovementioned cell and islet collectives could act as collective agents and have an agenda~\cite{levin2020cognition}, a job that has traditionally been reserved for vast neuronal networks in central nervous system. There is no particular reason why insulin-releasing cells would not aquire agency. First, on the molecular level, the spectrum of proteins pancreatic beta-cells express overlaps significantly with proteins expressed in neurons. Second, although a single islet harbors only an overseable number of a few hundred beta-cells organized as a sparse cell collective of an islet~\cite{Korosak2018FrontPhysiol}, the total number of beta-cells distributed in the pancreas can be compared to the number of photoreceptors in the retina, which is widely appreaciated as a sophisticated sensory system. It is quite possible that this agenda is encoded into intricate communication between the units as well as between the levels of distributed organisation of beta-cells, involving several biological levels of explanation~\cite{noble2012theory}.

What do we know about the function of beta-cell collectives so far? There are two specific features of beta-cell collective response to nutrient stimuli that have been well documented in the past: dependence of beta-cell activation, expressed as cytosolic Ca$^{2+}$ and membrane potential oscillations, on extracellular glucose concentration, and the biphasic nature of cytosolic Ca$^{2+}$ dynamics and insulin release.

The biphasic kinetics of insulin release following a supra-physiological rapid-onset and sustained glucose stimulation has been first described more than half a century ago in human~\cite{cerasi1967plasma}, rat~\cite{curry1968dynamics} and mouse pancreas~\cite{berglund1980different}. Ever since its discovery, several mechanisms have been proposed to describe this biphasity, ranging from cytosolic [Ca$^{2+}$]$_{c}$ handling from both intracellular~\cite{watras1991bell} and extracellular [Ca$^{2+}$] sources~\cite{wollheim1978roles}, insulin granule release probability~\cite{grodsky1972threshold,rorsman2003insulin,straub2004hypothesis}, inositol phosphate production~\cite{zawalich1988phosphoinositide}, and potentiating and inhibiting metabolic influences~\cite{nesher1987biphasic}.

Dual, Ca$^{2+}$ and metabolic oscillatory model has been previously used in an attempt to describe slow oscillations with a period between 5 to 10 minutes observed in isolated islets~\cite{Watts2014SIAM}. In more in situ preparation, like fresh pancreas tissue slices, a physiological stimulatory glucose concentration reproducibly triggers much faster Ca$^{2+}$ oscillatory activity with time scales too fast to support the involvement of the metabolic oscillators, however the biphasity is still evident~\cite{stovzer2021glucose}. This preparation helped us to uncover a new evidence regarding the function of intracellular Ca$^{2+}$ receptors underlying fast changes of the cytosolic Ca$^{2+}$ concentration in addition to previoulsy described sources~\cite{postic_intracellular_2021}. Based on these novel data, the concept congruent with all previously described mechanisms, including two independent mechanisms with sequent activation of IP$_3$ and ryanodine intracellular [Ca$^{2+}$] receptors, and incorporating several time scales of [Ca$^{2+}$]-induced-[Ca$^{2+}$] release (CICR) has been demonstrated~\cite{postic_intracellular_2021, Sluga2021Cells}. 

In this paper we model the empirically observed activation of beta-cell collectives with coupled stochastic oscillators and discuss how the combination of the abovementioned two intracellular Ca$^{2+}$ oscillatory modes fullfills all the criteria for an effective biological computation in determining the metabolic code. We suggest that particular spatial organization of the beta-cell agents on at least two levels of biological organization and the blood perfusion pattern render pancreas as an ultimate analytical and computational instrument for a succesful long-term metabolic economy. We show that to capture the whole spectrum of responses this type of biological computation needs to employ dual mode switching, sensing precision adjustment, response adaptivity and memory.

\begin{figure}
\centering
\includegraphics[width=0.95\linewidth]{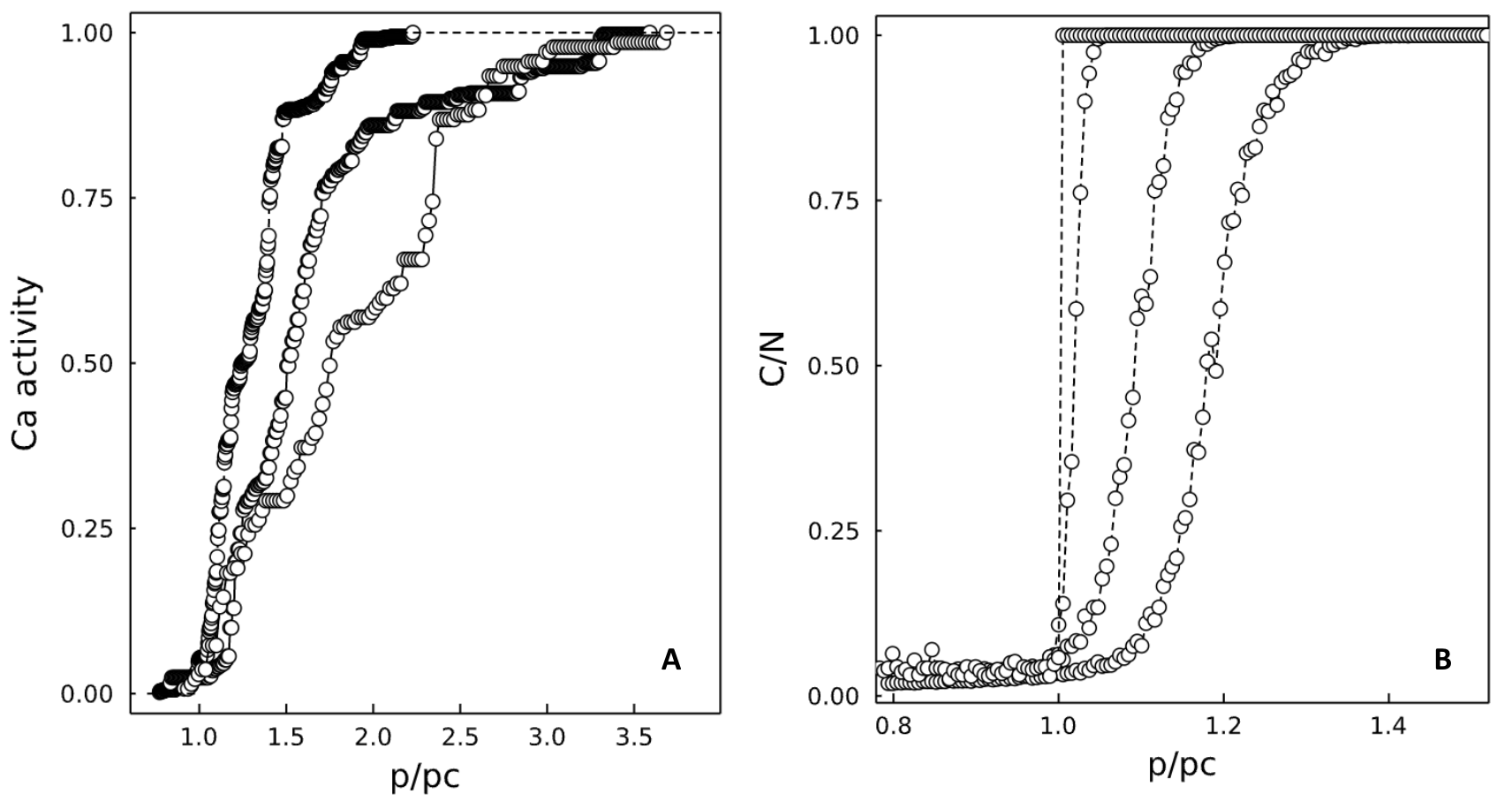}
\caption{\label{fig:fig1} {\bf Empirical data and model results.}
Glucose-dependent activity (A), from left to right: 16 mM, 12 mM, and 7 mM, data from~\cite{stovzer2021glucose}, size of giant component of correlation networks (B). Parameters used in model calculations: random regular network with $N=10^2$ and $k=10$, $T = 10^5$, 
$\Tilde\tau =2$, $\tau=\tau_{B1}=\tau_C = 10$. Noise $\Sigma = 0, 0.1, 3, 7$. Threshold value for correlation networks was set to $c_{0} = 0.1$. 
}
\end{figure}

\section{Model and results}
The typical response of the beta-cell collective activity at different glucose concentrations is shown in Figure 1A. Here, we present the beta-cell Ca$^{2+}$ activation as a function of a parameter $p$ defined as the time from the high glucose onset scaled with the half-time of activation (i.e. time interval in which half of cells activate) at the particular stimulus concentration. The response curves show two distinct behaviors with respect to glucose concentration: fast and abrupt transition to activity at high concetrations (16 mM and 12 mM in fig. 1A), and more gradual transition into the active state at lower glucose concentration (7 mM in fig. 1A).     

The presence of at least two distinct mechanisms with stochastic properties and the versatility of Ca$^{2+}$ release toolkit~\cite{watras1991bell,berridge1998calcium, berridge2000versatility, bootman2020fundamentals} allows us to start building models to better explain the activation of the beta-cell collectives.
We map the succession of Ca$^{2+}$ events onto the coupled stochastic oscillators model~\cite{nikitin2001collective} represented with a network where each node can be in one of three states $\sigma_i = \{\sigma_A, \sigma_B, \sigma_C\}$, and with the dynamics depending on the current state of the network $f\in [0,1]$ given by the fraction of nodes in the active state $\sigma_C$. Each node repeats the cycle: $\sigma_A\to \sigma_B\to \sigma_C\to \sigma_A$. 
Dynamics in state $\sigma_A$ is stochastic in the sense that period $\tau_A$, the time interval a node spends in $\sigma_A$, is randomly chosen from exponential distribution: $P(\tau_A) = \frac{1}{\tau}\exp(-\tau_A/\tau)$. 
State $\sigma_B$ is deterministic, but nodes in state $\sigma_B$ choose their waiting time depending on the state of the network and the threshold parameter $p$: they wait in state $\sigma_B$ with period $\tau_{B1}$ if $f<p$, or with $\tau_{B2}$ if $f>p$. Here $\tau_{B2} > \tau_{B1}$, so by switching to longer waiting times, the nodes that sense larger than threshold network state ($f>p$) tend to keep the network activity close to $p$. Nodes are active in the state $\sigma_C$ for time period $\tau_C$ in which they output a signal to all other nodes. 

Analysis and simulations of the original model~\cite{nikitin2001collective} showed that the phase of such all-to-all network depends on the value of the threshold parameter $p$. For very small values of $p$ almost all nodes stay in one mode and the collective network dynamics is stochastic with completely unsynchronised nodes activity. As $p$ approaches critical value $p_{c}$, the nodes start to switch between the two modes resulting in network transition into synchronised phase with oscillatory collective dynamics with period $\tau + \tau_{B2} + \tau_{C}$.

In our model here, we place $N$ cells on a random regular network where each node has $k$ nearest neighbors. The i-th cell senses the states of cells in $M_i$, the set of cells with the size $m_i = |M_i|$ that provide input to the i-th cell. We include the sensing precision of the cell that depends on the glucose concentration as the variance of $m_i$ at each time step. At high glucose concentrations (corresponding to 16 mM in fig.~1A) each cell has exactly $k$ nearest neighbors. At lower concentrations the number of neighbors fluctuates with $\Sigma$, a model parameter, that increases with diminishing glucose concentration. The rationale here is that as we approach the sensory threshold, beta-cells must increase their sensory precision which requires extending the communication with other cells in the islet. We model this neighborhood fluctuation by sampling additional number of neigbhors from normal distribution $\mathcal{N}(k,\Sigma)$ for each cell at each timestep. The cell compares its total input $f_i = (1/m_i)\sum_{j\in M_i}\delta(\sigma_j - \sigma_C)$ to the value of the threshold parameter $p$ and chooses the appropriate $\sigma_B$ state. The transition to coherent collective state at $p = p_{c}$ is possible only for $\Tilde\tau = \tau_{B2}/\tau_{B1} > 1$. We focus on the region of $p$ values where the network changes from stochastic to oscillatory collective dynamics. 

Following our previous approach where we transcripted the dynamics into correlation networks and studied their properties~\cite{podobnik2020beta,korosak2021}, 
we let the model run for $T$ timesteps at particular value of the control parameter $p$ to obtain $N$ timeseries ${x_i}$ of nodes activity. From these $N$ timeseries we constructed correlation networks of cells and follow the dependence of the size of the largest component $C$ on control parameter $p$. Starting with an empty network, for each pair of nodes $i,j$ we compute their cross-correlation $c_{ij} = \lra{x_i, x_j}$ and compare it to the chosen threshold value $c_{0}$. If $c_{ij} > c_0$ a link between $i$ and $j$ is placed in the network. Leaving the threshold $c_{0}$ fixed, the density of network increases with increasing $p$ along with activity coherence, and at particular critical value $p_c$ a phase transition in the size of largest cluster from $C/N\approx 0$ to $C/N\approx 1$ is observed (fig.~1B). Figure 1B shows the results of the model computations of largest component $C(p, \Sigma)/N$ as a function of model parameters $p$ and $\Sigma$. At low values of $\Sigma$ the response is abrupt, showing first-order like phase transition. With increasing $\Sigma$, the cells are able to explore their wider network neighborhood, sense more distant signals effectively increasing their interaction length and thus sensing precision~\cite{fancher2017fundamental}. This mode corresponds to the case of lower glucose concentrations where cells need to precisely determine the stimulus level and the response curve now resembles second-order like phase transition.    

\section{Discussion}
Our results show that we were able to reproduce main characteristics of beta-cell collective response to stimuli using a rather simple model of coupled fast stochastic oscillators mapped onto a correlation network. Two basic elements of the model are crucial to capture the whole spectrum of beta-cell response curves: (a) decision to switch between $\sigma_B$ states depending on the current collective state of the cell's neighbors, and (b) ability to control the sensing precision. Both elements together determine the abruptness of the disorder-order phase transition in collective beta-cell activity. Choosing between the two waiting times to keep the network state below a certain threshold results in a very abrupt, first-order like phase transition with the size of the giant component as a order parameter. This behavior is similar to an explosive percolation phenomenon in random network~\cite{achlioptas2009explosive}, where the growth of the giant component is delayed by competing links mechanism. The phase transition of beta-cell activation seem to change from first to second order like with gradual growth of activity, leading to the possibility that beta-cells adapt their collective behavior and move towards a critical point~\cite{mora2011biological} as the stimulus approaches the threshold value. These results suggest that further experiments with a slower rate of glucose exchange are required to capture a critical point and better determine the nature of phase transitions during the beta-cell activation.

\begin{figure}
\centering
\includegraphics[width=1.0\linewidth]{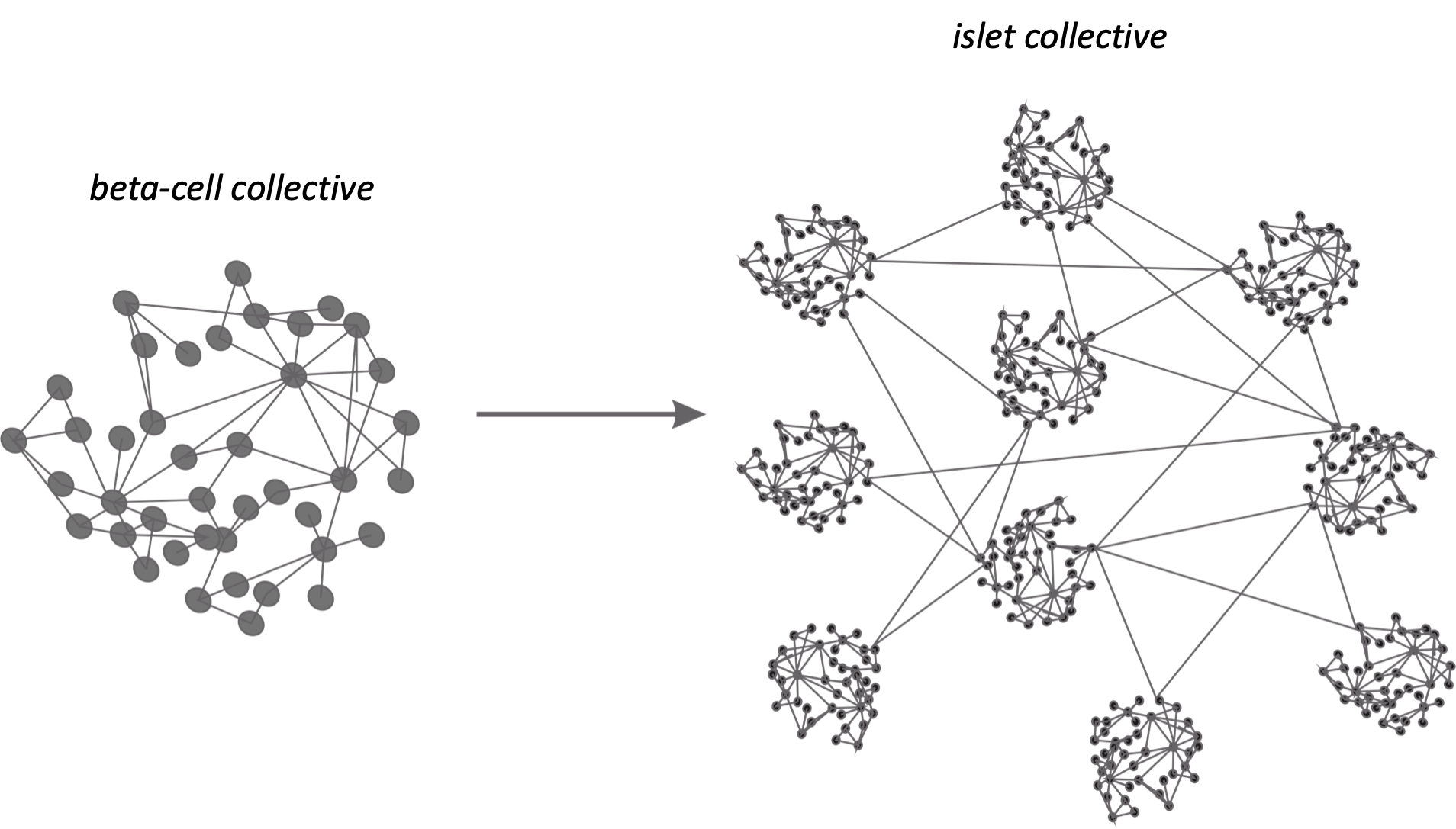}
\caption{\label{fig:fig2} {\bf Layers of fine-grained architecture in pancreas}.
}
\end{figure}

The activity of these fast stochastic oscillators in beta-cell or islet collective comes in a form of dynamics and probability distribution of Ca$^{2+}$ changes, which can be considered as information that fulfills all the requirements for computing. Unitary Ca$^{2+}$ events are triggered randomly, and get deterministically constrained into higher order spatial and temporal patterns with functional properties of the proteins involved. The information is communicated with temporal and spatial sampling through at least two layers of fine-grained architecture to ensure robustness, enhanced sensitivity or efficiency and evolvability enabling parallel terraces scans~\cite{672938}(fig.~2). We have previously described pancreatic islets, based on the number of interconnected beta-cells within such a collective and their average physical size, as sparse collectives with enhanced sensory function, able to resolve local gradients~\cite{Korosak2018FrontPhysiol}. In this way pancreas processes a certain volume of plasma at any given time, with dominant, immediate and focused responses to strong nutrient activators. At the same time a continous and unfocused exploration of weaker, but more diverse metabolic stimuli can be achieved, in a so called adaptive response.

At this stage we can try to readdress the question, why there are so many rather small islets in the pancreas, instead of just a single gland with a few hundred million beta-cells? Interestingly, the number of islets scales with the organism size, but not the size of the islets~\cite{JO20072655}. The distributed nature of pancreatic islets was reported to be required to locally support the function of acinar cells~\cite{henderson1969islets}. Alternative, both endocrine and exocrine pancreas could be part of the sensory system, exposed to plasma levels of nutrients and metabolites, computationally solving two different tasks. Endocrine cells are primarily responsible for the metabolic economy, while acinar cells with the release of digestive enzymes which modulate the intestinal digestion. Novel data on the exact pattern of pancreas microcirculation is required to reassess whether beta-cells can use local metabolite gradients to improve their computing function. Furthermore, since islets within pancreas are interconnected with neurons, another, higher level sparse collective of islets is a possibility, which could contribute to a detection of global metabolite concentration differences and in this way contribute to the real time analytical capacity of pancreas. Also here, new data on the pattern of pancreas macro- and microcirculation is required. The anatomical arrangement described above renders the distributed character of endocrine pancreas as a massive computer that parallelizes its computing both on the islet level as a computational unit and at least one higher level of biological organization.

\begin{figure}
\centering
\includegraphics[width=1.0\linewidth]{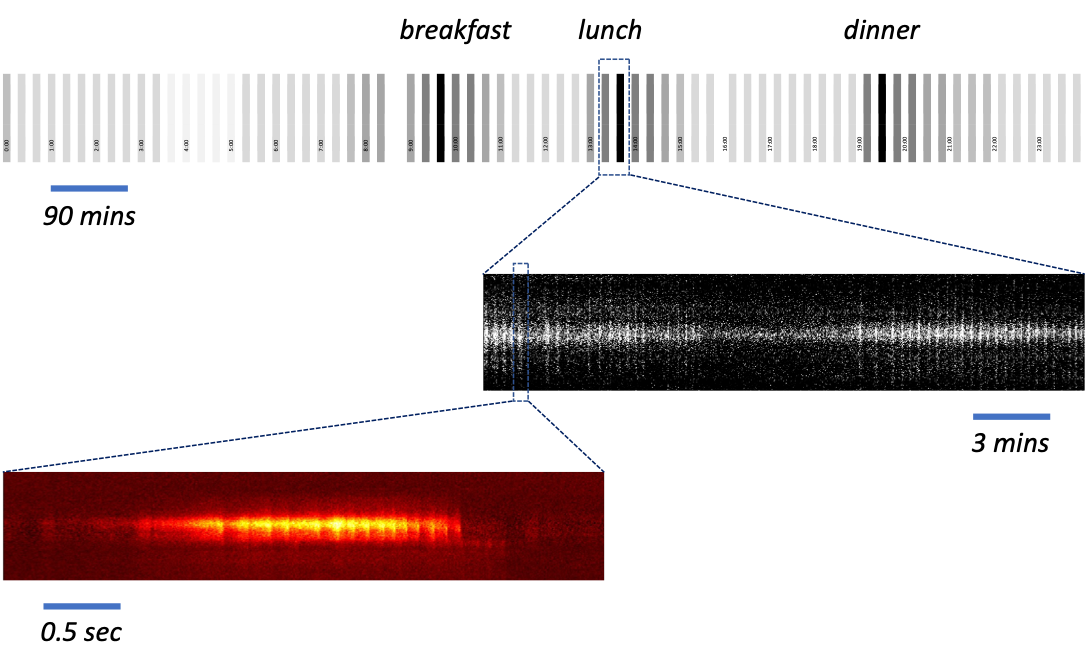}
\caption{\label{fig:fig3} {\bf Timescales of the collective activity in pancreas, measured as [Ca$^{2+}$]$_{c}$ events on a typical day}. Basal oscillatory activity with the period between 5-10 mins throughout the day (upper band) is increased after meals in a typical biphasic manner with a first peak followed by a plateau activity. On top of the basal oscillary activity, a faster series of few seconds long bursts develop (middle band) as a temporal sumation of unitary sub second events (lower band). The fast bursts can with higher stimulation further summate to slow events of some tens of seconds during the plateau phase. Both, IP$_3$ and ryanodine receptors contribute to the unitary events, however the former is dominant during the initial phase and the latter during the plateau phase.
}
\end{figure}

Similarly to the computation described for the immune system~\cite{mitchell2009complexity}, we suggest that pancreas has to combine two systems to compute whether the current metabolic challenge can be managed with small adjustments or treated as a threat which would justify a major hormonal response, but could in turn compromise the metabolic liquidity, i.e. instant availability of glucose for a vast majority of cells in an organism with obligatory glycolitic energy production requirement. Regular intervals between the meals as well as an initial and limited metabolic load can be intercepted by IP$_3$-receptor-dependent cytosolic Ca$^{2+}$ release in a form of regular slow oscillations. After a meal, however, a higher burden of different nutrients sparks the full biphasic response with initial slow component, followed by an adaptive RyR-dependent response (fig.~3). This latter activity can manifest itself in beta-cells as a spectrum of different phenotypes~\cite{manning2006oscillatory}. The main reason for this diversity of responses is that short sub second Ca$^{2+}$-induced Ca$^{2+}$ release events temporally summate into a variety of self-similar patterns of fast events that last for some seconds and slower events of some tens of seconds. This designed response is self-limited by reduced availability of Ca$^{2+}$ within the ER when nutrient levels drop postprandially~\cite{Klec_Cell_Physiol_Biochem_2019}. Oscillations of [Ca$^{2+}$]$_{c}$ result from the activity of IP$_3$ and ryanodine receptors, with the latter taking a more dominant role in later phases of stimulation or during stronger stimulations. Such duality of the contributions of these Ca$^{2+}$ release channels has been previously suggested to provide a basis for complex patterns of intracellular Ca$^{2+}$
regulation of neuronal activity~\cite{watras1991bell}.

An organisation like this would have further repercussions. For example, it can form a basis for a circuit memory, which could be amplified after a repetitive stimulation or which can be rewritten by stochastic editing to yield counterfactual future goals that would first express themselves only at the onset of a major perturbation, without leaving a clear causal link to the molecular organisational level, which is an exclusive focus of all current investigations~\cite{DURANT20172231}. And futhermore, a long-term, constant and strong activation of the adaptive system can lead to receptor destabilization and hyperactivity that can have potentially perilous effect on the health of endocrine pancreas and metabolic economy of an organism, eventually leading to diabetes mellitus. Recent experimental data support the last claim, since inhibition of the ryanodine receptor hyperactivity restored the dysfunctional glucose-induced [Ca$^{2+}$]$_{c}$ oscillations in the ER stress context~\cite{Yamamoto_JCB_2021}.

\begin{center}
--\,--\,--\,--\,--
\end{center}
\vspace{1mm}
\textbf{Acknowledgements.} MSR received financial support from NIH (R01DK127236) and from the Austrian Science Fund / Fonds zur F{\"o}rderung der Wissenschaftlichen Forschung (bilateral grants I3562--B27 and I4319--B30). AS, DK and MSR received financial support from the Slovenian Research Agency (research core funding program no.\ P3--0396 and projects no.\ N3-0048, no.\ N3-0133 and no.\ N3-9289). DK was further supported from the Slovenian Research Agency project no.\ J7-3156.

\noindent\textbf{Author contributions.} All authors contributed substantially to all aspects of the study.

\noindent\textbf{Conflict of interest.} The authors declare no conflict of interest, financial or otherwise.

\bibliography{refs}{}

\begin{thebibliography}{40}%
\makeatletter
\providecommand \@ifxundefined [1]{%
 \@ifx{#1\undefined}
}%
\providecommand \@ifnum [1]{%
 \ifnum #1\expandafter \@firstoftwo
 \else \expandafter \@secondoftwo
 \fi
}%
\providecommand \@ifx [1]{%
 \ifx #1\expandafter \@firstoftwo
 \else \expandafter \@secondoftwo
 \fi
}%
\providecommand \natexlab [1]{#1}%
\providecommand \enquote  [1]{``#1''}%
\providecommand \bibnamefont  [1]{#1}%
\providecommand \bibfnamefont [1]{#1}%
\providecommand \citenamefont [1]{#1}%
\providecommand \href@noop [0]{\@secondoftwo}%
\providecommand \href [0]{\begingroup \@sanitize@url \@href}%
\providecommand \@href[1]{\@@startlink{#1}\@@href}%
\providecommand \@@href[1]{\endgroup#1\@@endlink}%
\providecommand \@sanitize@url [0]{\catcode `\\12\catcode `\$12\catcode
  `\&12\catcode `\#12\catcode `\^12\catcode `\_12\catcode `\%12\relax}%
\providecommand \@@startlink[1]{}%
\providecommand \@@endlink[0]{}%
\providecommand \url  [0]{\begingroup\@sanitize@url \@url }%
\providecommand \@url [1]{\endgroup\@href {#1}{\urlprefix }}%
\providecommand \urlprefix  [0]{URL }%
\providecommand \Eprint [0]{\href }%
\providecommand \doibase [0]{http://dx.doi.org/}%
\providecommand \selectlanguage [0]{\@gobble}%
\providecommand \bibinfo  [0]{\@secondoftwo}%
\providecommand \bibfield  [0]{\@secondoftwo}%
\providecommand \translation [1]{[#1]}%
\providecommand \BibitemOpen [0]{}%
\providecommand \bibitemStop [0]{}%
\providecommand \bibitemNoStop [0]{.\EOS\space}%
\providecommand \EOS [0]{\spacefactor3000\relax}%
\providecommand \BibitemShut  [1]{\csname bibitem#1\endcsname}%
\let\auto@bib@innerbib\@empty
\bibitem [{\citenamefont {Mitchell}(2009)}]{mitchell2009complexity}%
  \BibitemOpen
  \bibfield  {author} {\bibinfo {author} {\bibfnamefont {M.}~\bibnamefont
  {Mitchell}},\ }\href@noop {} {\emph {\bibinfo {title} {Complexity: A guided
  tour}}}\ (\bibinfo  {publisher} {Oxford University Press},\ \bibinfo {year}
  {2009})\BibitemShut {NoStop}%
\bibitem [{\citenamefont {Tkacik}\ and\ \citenamefont
  {Bialek}(2016)}]{tkacik2016information}%
  \BibitemOpen
  \bibfield  {author} {\bibinfo {author} {\bibfnamefont {G.}~\bibnamefont
  {Tkacik}}\ and\ \bibinfo {author} {\bibfnamefont {W.}~\bibnamefont
  {Bialek}},\ }\href@noop {} {\bibfield  {journal} {\bibinfo  {journal} {The
  Annual Review of Condensed Matter Physics is}\ }\textbf {\bibinfo {volume}
  {7}},\ \bibinfo {pages} {12} (\bibinfo {year} {2016})}\BibitemShut {NoStop}%
\bibitem [{\citenamefont {Hopfield}(1994)}]{hopfield1994physics}%
  \BibitemOpen
  \bibfield  {author} {\bibinfo {author} {\bibfnamefont {J.}~\bibnamefont
  {Hopfield}},\ }\href@noop {} {\bibfield  {journal} {\bibinfo  {journal}
  {Journal of Theoretical Biology}\ }\textbf {\bibinfo {volume} {171}},\
  \bibinfo {pages} {53} (\bibinfo {year} {1994})}\BibitemShut {NoStop}%
\bibitem [{\citenamefont {Li}\ and\ \citenamefont
  {Bowerman}(2010)}]{li2010symmetry}%
  \BibitemOpen
  \bibfield  {author} {\bibinfo {author} {\bibfnamefont {R.}~\bibnamefont
  {Li}}\ and\ \bibinfo {author} {\bibfnamefont {B.}~\bibnamefont {Bowerman}},\
  }\href@noop {} {\bibfield  {journal} {\bibinfo  {journal} {Cold Spring Harbor
  perspectives in biology}\ }\textbf {\bibinfo {volume} {2}},\ \bibinfo {pages}
  {a003475} (\bibinfo {year} {2010})}\BibitemShut {NoStop}%
\bibitem [{\citenamefont {Mora}\ and\ \citenamefont
  {Bialek}(2011)}]{mora2011biological}%
  \BibitemOpen
  \bibfield  {author} {\bibinfo {author} {\bibfnamefont {T.}~\bibnamefont
  {Mora}}\ and\ \bibinfo {author} {\bibfnamefont {W.}~\bibnamefont {Bialek}},\
  }\href@noop {} {\bibfield  {journal} {\bibinfo  {journal} {Journal of
  Statistical Physics}\ }\textbf {\bibinfo {volume} {144}},\ \bibinfo {pages}
  {268} (\bibinfo {year} {2011})}\BibitemShut {NoStop}%
\bibitem [{\citenamefont {Langton}(1990)}]{langton1990computation}%
  \BibitemOpen
  \bibfield  {author} {\bibinfo {author} {\bibfnamefont {C.~G.}\ \bibnamefont
  {Langton}},\ }\href@noop {} {\bibfield  {journal} {\bibinfo  {journal}
  {Physica D: nonlinear phenomena}\ }\textbf {\bibinfo {volume} {42}},\
  \bibinfo {pages} {12} (\bibinfo {year} {1990})}\BibitemShut {NoStop}%
\bibitem [{\citenamefont {Cahill}(1970)}]{CahillNEJM1970}%
  \BibitemOpen
  \bibfield  {author} {\bibinfo {author} {\bibfnamefont {G.~F.}\ \bibnamefont
  {Cahill}},\ }\href {\doibase 10.1056/NEJM197003192821209} {\bibfield
  {journal} {\bibinfo  {journal} {New England Journal of Medicine}\ }\textbf
  {\bibinfo {volume} {282}},\ \bibinfo {pages} {668} (\bibinfo {year}
  {1970})},\ \bibinfo {note} {pMID: 4915800}\BibitemShut {NoStop}%
\bibitem [{\citenamefont {Haas}\ \emph {et~al.}(2022)\citenamefont {Haas},
  \citenamefont {Borsook}, \citenamefont {Adler},\ and\ \citenamefont
  {Freeman}}]{HAAS2022102983}%
  \BibitemOpen
  \bibfield  {author} {\bibinfo {author} {\bibfnamefont {A.}~\bibnamefont
  {Haas}}, \bibinfo {author} {\bibfnamefont {D.}~\bibnamefont {Borsook}},
  \bibinfo {author} {\bibfnamefont {G.}~\bibnamefont {Adler}}, \ and\ \bibinfo
  {author} {\bibfnamefont {R.}~\bibnamefont {Freeman}},\ }\href {\doibase
  https://doi.org/10.1016/j.autneu.2022.102983} {\bibfield  {journal} {\bibinfo
   {journal} {Autonomic Neuroscience}\ }\textbf {\bibinfo {volume} {240}},\
  \bibinfo {pages} {102983} (\bibinfo {year} {2022})}\BibitemShut {NoStop}%
\bibitem [{\citenamefont {Levin}\ and\ \citenamefont
  {Dennett}(2020)}]{levin2020cognition}%
  \BibitemOpen
  \bibfield  {author} {\bibinfo {author} {\bibfnamefont {M.}~\bibnamefont
  {Levin}}\ and\ \bibinfo {author} {\bibfnamefont {D.~C.}\ \bibnamefont
  {Dennett}},\ }\href@noop {} {\bibfield  {journal} {\bibinfo  {journal} {Aeon
  Essays. Retrieved}\ } (\bibinfo {year} {2020})}\BibitemShut {NoStop}%
\bibitem [{\citenamefont {Korošak}\ and\ \citenamefont
  {Slak~Rupnik}(2018)}]{Korosak2018FrontPhysiol}%
  \BibitemOpen
  \bibfield  {author} {\bibinfo {author} {\bibfnamefont {D.}~\bibnamefont
  {Korošak}}\ and\ \bibinfo {author} {\bibfnamefont {M.}~\bibnamefont
  {Slak~Rupnik}},\ }\href {\doibase 10.3389/fphys.2018.00031} {\bibfield
  {journal} {\bibinfo  {journal} {Frontiers in Physiology}\ }\textbf {\bibinfo
  {volume} {9}} (\bibinfo {year} {2018}),\
  10.3389/fphys.2018.00031}\BibitemShut {NoStop}%
\bibitem [{\citenamefont {Noble}(2012)}]{noble2012theory}%
  \BibitemOpen
  \bibfield  {author} {\bibinfo {author} {\bibfnamefont {D.}~\bibnamefont
  {Noble}},\ }\href@noop {} {\bibfield  {journal} {\bibinfo  {journal}
  {Interface focus}\ }\textbf {\bibinfo {volume} {2}},\ \bibinfo {pages} {55}
  (\bibinfo {year} {2012})}\BibitemShut {NoStop}%
\bibitem [{\citenamefont {Cerasi}\ and\ \citenamefont
  {Luft}(1967)}]{cerasi1967plasma}%
  \BibitemOpen
  \bibfield  {author} {\bibinfo {author} {\bibfnamefont {E.}~\bibnamefont
  {Cerasi}}\ and\ \bibinfo {author} {\bibfnamefont {R.}~\bibnamefont {Luft}},\
  }\href@noop {} {\bibfield  {journal} {\bibinfo  {journal} {European Journal
  of Endocrinology}\ }\textbf {\bibinfo {volume} {55}},\ \bibinfo {pages} {278}
  (\bibinfo {year} {1967})}\BibitemShut {NoStop}%
\bibitem [{\citenamefont {Curry}\ \emph {et~al.}(1968)\citenamefont {Curry},
  \citenamefont {Bennett},\ and\ \citenamefont {Grodsky}}]{curry1968dynamics}%
  \BibitemOpen
  \bibfield  {author} {\bibinfo {author} {\bibfnamefont {D.~L.}\ \bibnamefont
  {Curry}}, \bibinfo {author} {\bibfnamefont {L.~L.}\ \bibnamefont {Bennett}},
  \ and\ \bibinfo {author} {\bibfnamefont {G.~M.}\ \bibnamefont {Grodsky}},\
  }\href@noop {} {\bibfield  {journal} {\bibinfo  {journal} {Endocrinology}\
  }\textbf {\bibinfo {volume} {83}},\ \bibinfo {pages} {572} (\bibinfo {year}
  {1968})}\BibitemShut {NoStop}%
\bibitem [{\citenamefont {Berglund}(1980)}]{berglund1980different}%
  \BibitemOpen
  \bibfield  {author} {\bibinfo {author} {\bibfnamefont {O.}~\bibnamefont
  {Berglund}},\ }\href@noop {} {\bibfield  {journal} {\bibinfo  {journal}
  {European Journal of Endocrinology}\ }\textbf {\bibinfo {volume} {93}},\
  \bibinfo {pages} {54} (\bibinfo {year} {1980})}\BibitemShut {NoStop}%
\bibitem [{\citenamefont {Watras}\ \emph {et~al.}(1991)\citenamefont {Watras},
  \citenamefont {Ehrlich} \emph {et~al.}}]{watras1991bell}%
  \BibitemOpen
  \bibfield  {author} {\bibinfo {author} {\bibfnamefont {J.}~\bibnamefont
  {Watras}}, \bibinfo {author} {\bibfnamefont {B.~E.}\ \bibnamefont {Ehrlich}},
   \emph {et~al.},\ }\href@noop {} {\bibfield  {journal} {\bibinfo  {journal}
  {Nature}\ }\textbf {\bibinfo {volume} {351}},\ \bibinfo {pages} {751}
  (\bibinfo {year} {1991})}\BibitemShut {NoStop}%
\bibitem [{\citenamefont {Wollheim}\ \emph {et~al.}(1978)\citenamefont
  {Wollheim}, \citenamefont {Kikuchi}, \citenamefont {Renold}, \citenamefont
  {Sharp} \emph {et~al.}}]{wollheim1978roles}%
  \BibitemOpen
  \bibfield  {author} {\bibinfo {author} {\bibfnamefont {C.~B.}\ \bibnamefont
  {Wollheim}}, \bibinfo {author} {\bibfnamefont {M.}~\bibnamefont {Kikuchi}},
  \bibinfo {author} {\bibfnamefont {A.~E.}\ \bibnamefont {Renold}}, \bibinfo
  {author} {\bibfnamefont {G.~W.}\ \bibnamefont {Sharp}},  \emph {et~al.},\
  }\href@noop {} {\bibfield  {journal} {\bibinfo  {journal} {The Journal of
  clinical investigation}\ }\textbf {\bibinfo {volume} {62}},\ \bibinfo {pages}
  {451} (\bibinfo {year} {1978})}\BibitemShut {NoStop}%
\bibitem [{\citenamefont {Grodsky}\ \emph {et~al.}(1972)\citenamefont {Grodsky}
  \emph {et~al.}}]{grodsky1972threshold}%
  \BibitemOpen
  \bibfield  {author} {\bibinfo {author} {\bibfnamefont {G.~M.}\ \bibnamefont
  {Grodsky}} \emph {et~al.},\ }\href@noop {} {\bibfield  {journal} {\bibinfo
  {journal} {The Journal of clinical investigation}\ }\textbf {\bibinfo
  {volume} {51}},\ \bibinfo {pages} {2047} (\bibinfo {year}
  {1972})}\BibitemShut {NoStop}%
\bibitem [{\citenamefont {Rorsman}\ and\ \citenamefont
  {Renstr{\"o}m}(2003)}]{rorsman2003insulin}%
  \BibitemOpen
  \bibfield  {author} {\bibinfo {author} {\bibfnamefont {P.}~\bibnamefont
  {Rorsman}}\ and\ \bibinfo {author} {\bibfnamefont {E.}~\bibnamefont
  {Renstr{\"o}m}},\ }\href@noop {} {\bibfield  {journal} {\bibinfo  {journal}
  {Diabetologia}\ }\textbf {\bibinfo {volume} {46}},\ \bibinfo {pages} {1029}
  (\bibinfo {year} {2003})}\BibitemShut {NoStop}%
\bibitem [{\citenamefont {Straub}\ and\ \citenamefont
  {Sharp}(2004)}]{straub2004hypothesis}%
  \BibitemOpen
  \bibfield  {author} {\bibinfo {author} {\bibfnamefont {S.~G.}\ \bibnamefont
  {Straub}}\ and\ \bibinfo {author} {\bibfnamefont {G.~W.}\ \bibnamefont
  {Sharp}},\ }\href@noop {} {\bibfield  {journal} {\bibinfo  {journal}
  {American Journal of Physiology-Cell Physiology}\ }\textbf {\bibinfo {volume}
  {287}},\ \bibinfo {pages} {C565} (\bibinfo {year} {2004})}\BibitemShut
  {NoStop}%
\bibitem [{\citenamefont {Zawalich}\ and\ \citenamefont
  {Zawalich}(1988)}]{zawalich1988phosphoinositide}%
  \BibitemOpen
  \bibfield  {author} {\bibinfo {author} {\bibfnamefont {W.~S.}\ \bibnamefont
  {Zawalich}}\ and\ \bibinfo {author} {\bibfnamefont {K.~C.}\ \bibnamefont
  {Zawalich}},\ }\href@noop {} {\bibfield  {journal} {\bibinfo  {journal}
  {Diabetes}\ }\textbf {\bibinfo {volume} {37}},\ \bibinfo {pages} {1294}
  (\bibinfo {year} {1988})}\BibitemShut {NoStop}%
\bibitem [{\citenamefont {Nesher}\ and\ \citenamefont
  {Cerasi}(1987)}]{nesher1987biphasic}%
  \BibitemOpen
  \bibfield  {author} {\bibinfo {author} {\bibfnamefont {R.}~\bibnamefont
  {Nesher}}\ and\ \bibinfo {author} {\bibfnamefont {E.}~\bibnamefont
  {Cerasi}},\ }\href@noop {} {\bibfield  {journal} {\bibinfo  {journal}
  {Endocrinology}\ }\textbf {\bibinfo {volume} {121}},\ \bibinfo {pages} {1017}
  (\bibinfo {year} {1987})}\BibitemShut {NoStop}%
\bibitem [{\citenamefont {Watts}\ \emph {et~al.}(2014)\citenamefont {Watts},
  \citenamefont {Fendler}, \citenamefont {Merrins}, \citenamefont {Satin},
  \citenamefont {Bertram},\ and\ \citenamefont {Sherman}}]{Watts2014SIAM}%
  \BibitemOpen
  \bibfield  {author} {\bibinfo {author} {\bibfnamefont {M.}~\bibnamefont
  {Watts}}, \bibinfo {author} {\bibfnamefont {B.}~\bibnamefont {Fendler}},
  \bibinfo {author} {\bibfnamefont {M.~J.}\ \bibnamefont {Merrins}}, \bibinfo
  {author} {\bibfnamefont {L.~S.}\ \bibnamefont {Satin}}, \bibinfo {author}
  {\bibfnamefont {R.}~\bibnamefont {Bertram}}, \ and\ \bibinfo {author}
  {\bibfnamefont {A.}~\bibnamefont {Sherman}},\ }\href {\doibase
  10.1137/130920198} {\bibfield  {journal} {\bibinfo  {journal} {SIAM J Appl
  Dyn Syst}\ }\textbf {\bibinfo {volume} {13}},\ \bibinfo {pages} {683}
  (\bibinfo {year} {2014})}\BibitemShut {NoStop}%
\bibitem [{\citenamefont {Sto{\v{z}}er}\ \emph {et~al.}(2021)\citenamefont
  {Sto{\v{z}}er}, \citenamefont {Skelin~Klemen}, \citenamefont {Gosak},
  \citenamefont {Kri{\v{z}}an{\v{c}}i{\'c}~Bombek}, \citenamefont {Pohorec},
  \citenamefont {Slak~Rupnik},\ and\ \citenamefont
  {Dolen{\v{s}}ek}}]{stovzer2021glucose}%
  \BibitemOpen
  \bibfield  {author} {\bibinfo {author} {\bibfnamefont {A.}~\bibnamefont
  {Sto{\v{z}}er}}, \bibinfo {author} {\bibfnamefont {M.}~\bibnamefont
  {Skelin~Klemen}}, \bibinfo {author} {\bibfnamefont {M.}~\bibnamefont
  {Gosak}}, \bibinfo {author} {\bibfnamefont {L.}~\bibnamefont
  {Kri{\v{z}}an{\v{c}}i{\'c}~Bombek}}, \bibinfo {author} {\bibfnamefont
  {V.}~\bibnamefont {Pohorec}}, \bibinfo {author} {\bibfnamefont
  {M.}~\bibnamefont {Slak~Rupnik}}, \ and\ \bibinfo {author} {\bibfnamefont
  {J.}~\bibnamefont {Dolen{\v{s}}ek}},\ }\href@noop {} {\bibfield  {journal}
  {\bibinfo  {journal} {American Journal of Physiology-Endocrinology and
  Metabolism}\ }\textbf {\bibinfo {volume} {321}},\ \bibinfo {pages} {E305}
  (\bibinfo {year} {2021})}\BibitemShut {NoStop}%
\bibitem [{\citenamefont {Posti{\'c}}\ \emph {et~al.}(2021)\citenamefont
  {Posti{\'c}}, \citenamefont {Sarikas}, \citenamefont {Pfabe}, \citenamefont
  {Pohorec}, \citenamefont {Kri{\v z}an{\v c}i{\'c}~Bombek}, \citenamefont
  {Sluga}, \citenamefont {Skelin~Klemen}, \citenamefont {Dolen{\v s}ek},
  \citenamefont {Koro{\v s}ak}, \citenamefont {Sto{\v z}er}, \citenamefont
  {Evans-Molina}, \citenamefont {Johnson},\ and\ \citenamefont
  {Slak~Rupnik}}]{postic_intracellular_2021}%
  \BibitemOpen
  \bibfield  {author} {\bibinfo {author} {\bibfnamefont {S.}~\bibnamefont
  {Posti{\'c}}}, \bibinfo {author} {\bibfnamefont {S.}~\bibnamefont {Sarikas}},
  \bibinfo {author} {\bibfnamefont {J.}~\bibnamefont {Pfabe}}, \bibinfo
  {author} {\bibfnamefont {V.}~\bibnamefont {Pohorec}}, \bibinfo {author}
  {\bibfnamefont {L.}~\bibnamefont {Kri{\v z}an{\v c}i{\'c}~Bombek}}, \bibinfo
  {author} {\bibfnamefont {N.}~\bibnamefont {Sluga}}, \bibinfo {author}
  {\bibfnamefont {M.}~\bibnamefont {Skelin~Klemen}}, \bibinfo {author}
  {\bibfnamefont {J.}~\bibnamefont {Dolen{\v s}ek}}, \bibinfo {author}
  {\bibfnamefont {D.}~\bibnamefont {Koro{\v s}ak}}, \bibinfo {author}
  {\bibfnamefont {A.}~\bibnamefont {Sto{\v z}er}}, \bibinfo {author}
  {\bibfnamefont {C.}~\bibnamefont {Evans-Molina}}, \bibinfo {author}
  {\bibfnamefont {J.~D.}\ \bibnamefont {Johnson}}, \ and\ \bibinfo {author}
  {\bibfnamefont {M.}~\bibnamefont {Slak~Rupnik}},\ }\href {\doibase
  10.1101/2021.04.14.439796} {\bibfield  {journal} {\bibinfo  {journal}
  {bioRxiv}\ } (\bibinfo {year} {2021}),\
  10.1101/2021.04.14.439796}\BibitemShut {NoStop}%
\bibitem [{\citenamefont {Sluga}\ \emph {et~al.}(2021)\citenamefont {Sluga},
  \citenamefont {Postić}, \citenamefont {Sarikas}, \citenamefont {Huang},
  \citenamefont {Stožer},\ and\ \citenamefont {Slak~Rupnik}}]{Sluga2021Cells}%
  \BibitemOpen
  \bibfield  {author} {\bibinfo {author} {\bibfnamefont {N.}~\bibnamefont
  {Sluga}}, \bibinfo {author} {\bibfnamefont {S.}~\bibnamefont {Postić}},
  \bibinfo {author} {\bibfnamefont {S.}~\bibnamefont {Sarikas}}, \bibinfo
  {author} {\bibfnamefont {Y.-C.}\ \bibnamefont {Huang}}, \bibinfo {author}
  {\bibfnamefont {A.}~\bibnamefont {Stožer}}, \ and\ \bibinfo {author}
  {\bibfnamefont {M.}~\bibnamefont {Slak~Rupnik}},\ }\href {\doibase
  10.3390/cells10071580} {\bibfield  {journal} {\bibinfo  {journal} {Cells}\
  }\textbf {\bibinfo {volume} {10}},\ \bibinfo {pages} {1580} (\bibinfo {year}
  {2021})}\BibitemShut {NoStop}%
\bibitem [{\citenamefont {Berridge}\ \emph {et~al.}(1998)\citenamefont
  {Berridge}, \citenamefont {Bootman},\ and\ \citenamefont
  {Lipp}}]{berridge1998calcium}%
  \BibitemOpen
  \bibfield  {author} {\bibinfo {author} {\bibfnamefont {M.~J.}\ \bibnamefont
  {Berridge}}, \bibinfo {author} {\bibfnamefont {M.~D.}\ \bibnamefont
  {Bootman}}, \ and\ \bibinfo {author} {\bibfnamefont {P.}~\bibnamefont
  {Lipp}},\ }\href@noop {} {\bibfield  {journal} {\bibinfo  {journal} {Nature}\
  }\textbf {\bibinfo {volume} {395}},\ \bibinfo {pages} {645} (\bibinfo {year}
  {1998})}\BibitemShut {NoStop}%
\bibitem [{\citenamefont {Berridge}\ \emph {et~al.}(2000)\citenamefont
  {Berridge}, \citenamefont {Lipp},\ and\ \citenamefont
  {Bootman}}]{berridge2000versatility}%
  \BibitemOpen
  \bibfield  {author} {\bibinfo {author} {\bibfnamefont {M.~J.}\ \bibnamefont
  {Berridge}}, \bibinfo {author} {\bibfnamefont {P.}~\bibnamefont {Lipp}}, \
  and\ \bibinfo {author} {\bibfnamefont {M.~D.}\ \bibnamefont {Bootman}},\
  }\href@noop {} {\bibfield  {journal} {\bibinfo  {journal} {Nature reviews
  Molecular cell biology}\ }\textbf {\bibinfo {volume} {1}},\ \bibinfo {pages}
  {11} (\bibinfo {year} {2000})}\BibitemShut {NoStop}%
\bibitem [{\citenamefont {Bootman}\ and\ \citenamefont
  {Bultynck}(2020)}]{bootman2020fundamentals}%
  \BibitemOpen
  \bibfield  {author} {\bibinfo {author} {\bibfnamefont {M.~D.}\ \bibnamefont
  {Bootman}}\ and\ \bibinfo {author} {\bibfnamefont {G.}~\bibnamefont
  {Bultynck}},\ }\href@noop {} {\bibfield  {journal} {\bibinfo  {journal} {Cold
  Spring Harbor perspectives in biology}\ }\textbf {\bibinfo {volume} {12}},\
  \bibinfo {pages} {a038802} (\bibinfo {year} {2020})}\BibitemShut {NoStop}%
\bibitem [{\citenamefont {Nikitin}\ \emph {et~al.}(2001)\citenamefont
  {Nikitin}, \citenamefont {N{\'e}da},\ and\ \citenamefont
  {Vicsek}}]{nikitin2001collective}%
  \BibitemOpen
  \bibfield  {author} {\bibinfo {author} {\bibfnamefont {A.}~\bibnamefont
  {Nikitin}}, \bibinfo {author} {\bibfnamefont {Z.}~\bibnamefont {N{\'e}da}}, \
  and\ \bibinfo {author} {\bibfnamefont {T.}~\bibnamefont {Vicsek}},\
  }\href@noop {} {\bibfield  {journal} {\bibinfo  {journal} {Physical Review
  Letters}\ }\textbf {\bibinfo {volume} {87}},\ \bibinfo {pages} {024101}
  (\bibinfo {year} {2001})}\BibitemShut {NoStop}%
\bibitem [{\citenamefont {Podobnik}\ \emph {et~al.}(2020)\citenamefont
  {Podobnik}, \citenamefont {Koro{\v{s}}ak}, \citenamefont {Klemen},
  \citenamefont {Sto{\v{z}}er}, \citenamefont {Dolen{\v{s}}ek}, \citenamefont
  {Rupnik}, \citenamefont {Ivanov}, \citenamefont {Holme},\ and\ \citenamefont
  {Jusup}}]{podobnik2020beta}%
  \BibitemOpen
  \bibfield  {author} {\bibinfo {author} {\bibfnamefont {B.}~\bibnamefont
  {Podobnik}}, \bibinfo {author} {\bibfnamefont {D.}~\bibnamefont
  {Koro{\v{s}}ak}}, \bibinfo {author} {\bibfnamefont {M.~S.}\ \bibnamefont
  {Klemen}}, \bibinfo {author} {\bibfnamefont {A.}~\bibnamefont
  {Sto{\v{z}}er}}, \bibinfo {author} {\bibfnamefont {J.}~\bibnamefont
  {Dolen{\v{s}}ek}}, \bibinfo {author} {\bibfnamefont {M.~S.}\ \bibnamefont
  {Rupnik}}, \bibinfo {author} {\bibfnamefont {P.~C.}\ \bibnamefont {Ivanov}},
  \bibinfo {author} {\bibfnamefont {P.}~\bibnamefont {Holme}}, \ and\ \bibinfo
  {author} {\bibfnamefont {M.}~\bibnamefont {Jusup}},\ }\href@noop {}
  {\bibfield  {journal} {\bibinfo  {journal} {Biophysical journal}\ }\textbf
  {\bibinfo {volume} {118}},\ \bibinfo {pages} {2588} (\bibinfo {year}
  {2020})}\BibitemShut {NoStop}%
\bibitem [{\citenamefont {Koro\v{s}ak}\ \emph {et~al.}(2021)\citenamefont
  {Koro\v{s}ak}, \citenamefont {Jusup}, \citenamefont {Podobnik}, \citenamefont
  {Sto\v{z}er}, \citenamefont {Dolen\v{s}ek}, \citenamefont {Holme},\ and\
  \citenamefont {Rupnik}}]{korosak2021}%
  \BibitemOpen
  \bibfield  {author} {\bibinfo {author} {\bibfnamefont {D.}~\bibnamefont
  {Koro\v{s}ak}}, \bibinfo {author} {\bibfnamefont {M.}~\bibnamefont {Jusup}},
  \bibinfo {author} {\bibfnamefont {B.}~\bibnamefont {Podobnik}}, \bibinfo
  {author} {\bibfnamefont {A.}~\bibnamefont {Sto\v{z}er}}, \bibinfo {author}
  {\bibfnamefont {J.}~\bibnamefont {Dolen\v{s}ek}}, \bibinfo {author}
  {\bibfnamefont {P.}~\bibnamefont {Holme}}, \ and\ \bibinfo {author}
  {\bibfnamefont {M.~S.}\ \bibnamefont {Rupnik}},\ }\href {\doibase
  10.1103/PhysRevLett.127.168101} {\bibfield  {journal} {\bibinfo  {journal}
  {Phys. Rev. Lett.}\ }\textbf {\bibinfo {volume} {127}},\ \bibinfo {pages}
  {168101} (\bibinfo {year} {2021})}\BibitemShut {NoStop}%
\bibitem [{\citenamefont {Fancher}\ and\ \citenamefont
  {Mugler}(2017)}]{fancher2017fundamental}%
  \BibitemOpen
  \bibfield  {author} {\bibinfo {author} {\bibfnamefont {S.}~\bibnamefont
  {Fancher}}\ and\ \bibinfo {author} {\bibfnamefont {A.}~\bibnamefont
  {Mugler}},\ }\href@noop {} {\bibfield  {journal} {\bibinfo  {journal}
  {Physical review letters}\ }\textbf {\bibinfo {volume} {118}},\ \bibinfo
  {pages} {078101} (\bibinfo {year} {2017})}\BibitemShut {NoStop}%
\bibitem [{\citenamefont {Achlioptas}\ \emph {et~al.}(2009)\citenamefont
  {Achlioptas}, \citenamefont {D'Souza},\ and\ \citenamefont
  {Spencer}}]{achlioptas2009explosive}%
  \BibitemOpen
  \bibfield  {author} {\bibinfo {author} {\bibfnamefont {D.}~\bibnamefont
  {Achlioptas}}, \bibinfo {author} {\bibfnamefont {R.~M.}\ \bibnamefont
  {D'Souza}}, \ and\ \bibinfo {author} {\bibfnamefont {J.}~\bibnamefont
  {Spencer}},\ }\href@noop {} {\bibfield  {journal} {\bibinfo  {journal}
  {Science}\ }\textbf {\bibinfo {volume} {323}},\ \bibinfo {pages} {1453}
  (\bibinfo {year} {2009})}\BibitemShut {NoStop}%
\bibitem [{\citenamefont {Rehling}\ and\ \citenamefont
  {Hofstadter}(1997)}]{672938}%
  \BibitemOpen
  \bibfield  {author} {\bibinfo {author} {\bibfnamefont {J.}~\bibnamefont
  {Rehling}}\ and\ \bibinfo {author} {\bibfnamefont {D.}~\bibnamefont
  {Hofstadter}},\ }in\ \href {\doibase 10.1109/ICIPS.1997.672938} {\emph
  {\bibinfo {booktitle} {1997 IEEE International Conference on Intelligent
  Processing Systems (Cat. No.97TH8335)}}},\ Vol.~\bibinfo {volume} {1}\
  (\bibinfo {year} {1997})\ pp.\ \bibinfo {pages} {900--904 vol.1}\BibitemShut
  {NoStop}%
\bibitem [{\citenamefont {Jo}\ \emph {et~al.}(2007)\citenamefont {Jo},
  \citenamefont {Choi},\ and\ \citenamefont {Koh}}]{JO20072655}%
  \BibitemOpen
  \bibfield  {author} {\bibinfo {author} {\bibfnamefont {J.}~\bibnamefont
  {Jo}}, \bibinfo {author} {\bibfnamefont {M.~Y.}\ \bibnamefont {Choi}}, \ and\
  \bibinfo {author} {\bibfnamefont {D.-S.}\ \bibnamefont {Koh}},\ }\href
  {\doibase https://doi.org/10.1529/biophysj.107.104125} {\bibfield  {journal}
  {\bibinfo  {journal} {Biophysical Journal}\ }\textbf {\bibinfo {volume}
  {93}},\ \bibinfo {pages} {2655} (\bibinfo {year} {2007})}\BibitemShut
  {NoStop}%
\bibitem [{\citenamefont {Henderson}(1969)}]{henderson1969islets}%
  \BibitemOpen
  \bibfield  {author} {\bibinfo {author} {\bibfnamefont {J.}~\bibnamefont
  {Henderson}},\ }\href@noop {} {\bibfield  {journal} {\bibinfo  {journal} {The
  Lancet}\ }\textbf {\bibinfo {volume} {294}},\ \bibinfo {pages} {469}
  (\bibinfo {year} {1969})}\BibitemShut {NoStop}%
\bibitem [{\citenamefont {Manning~Fox}\ \emph {et~al.}(2006)\citenamefont
  {Manning~Fox}, \citenamefont {Gyulkhandanyan}, \citenamefont {Satin},\ and\
  \citenamefont {Wheeler}}]{manning2006oscillatory}%
  \BibitemOpen
  \bibfield  {author} {\bibinfo {author} {\bibfnamefont {J.~E.}\ \bibnamefont
  {Manning~Fox}}, \bibinfo {author} {\bibfnamefont {A.~V.}\ \bibnamefont
  {Gyulkhandanyan}}, \bibinfo {author} {\bibfnamefont {L.~S.}\ \bibnamefont
  {Satin}}, \ and\ \bibinfo {author} {\bibfnamefont {M.~B.}\ \bibnamefont
  {Wheeler}},\ }\href@noop {} {\bibfield  {journal} {\bibinfo  {journal}
  {Endocrinology}\ }\textbf {\bibinfo {volume} {147}},\ \bibinfo {pages} {4655}
  (\bibinfo {year} {2006})}\BibitemShut {NoStop}%
\bibitem [{\citenamefont {Klec}\ \emph {et~al.}(2019)\citenamefont {Klec},
  \citenamefont {Madreiter-Sokolowski}, \citenamefont {Stryeck}, \citenamefont
  {Sachdev}, \citenamefont {Duta-Mare}, \citenamefont {Gottschalk},
  \citenamefont {Depaoli}, \citenamefont {Rost}, \citenamefont {Hay},
  \citenamefont {Waldeck-Weiermair}, \citenamefont {Kratky}, \citenamefont
  {Madl}, \citenamefont {Malli},\ and\ \citenamefont
  {Graier}}]{Klec_Cell_Physiol_Biochem_2019}%
  \BibitemOpen
  \bibfield  {author} {\bibinfo {author} {\bibfnamefont {C.}~\bibnamefont
  {Klec}}, \bibinfo {author} {\bibfnamefont {C.~T.}\ \bibnamefont
  {Madreiter-Sokolowski}}, \bibinfo {author} {\bibfnamefont {S.}~\bibnamefont
  {Stryeck}}, \bibinfo {author} {\bibfnamefont {V.}~\bibnamefont {Sachdev}},
  \bibinfo {author} {\bibfnamefont {M.}~\bibnamefont {Duta-Mare}}, \bibinfo
  {author} {\bibfnamefont {B.}~\bibnamefont {Gottschalk}}, \bibinfo {author}
  {\bibfnamefont {M.~R.}\ \bibnamefont {Depaoli}}, \bibinfo {author}
  {\bibfnamefont {R.}~\bibnamefont {Rost}}, \bibinfo {author} {\bibfnamefont
  {J.}~\bibnamefont {Hay}}, \bibinfo {author} {\bibfnamefont {M.}~\bibnamefont
  {Waldeck-Weiermair}}, \bibinfo {author} {\bibfnamefont {D.}~\bibnamefont
  {Kratky}}, \bibinfo {author} {\bibfnamefont {T.}~\bibnamefont {Madl}},
  \bibinfo {author} {\bibfnamefont {R.}~\bibnamefont {Malli}}, \ and\ \bibinfo
  {author} {\bibfnamefont {W.~F.}\ \bibnamefont {Graier}},\ }\href {\doibase
  10.33594/000000005} {\bibfield  {journal} {\bibinfo  {journal} {Cell Physiol
  Biochem}\ }\textbf {\bibinfo {volume} {52}},\ \bibinfo {pages} {57} (\bibinfo
  {year} {2019})}\BibitemShut {NoStop}%
\bibitem [{\citenamefont {Durant}\ \emph {et~al.}(2017)\citenamefont {Durant},
  \citenamefont {Morokuma}, \citenamefont {Fields}, \citenamefont {Williams},
  \citenamefont {Adams},\ and\ \citenamefont {Levin}}]{DURANT20172231}%
  \BibitemOpen
  \bibfield  {author} {\bibinfo {author} {\bibfnamefont {F.}~\bibnamefont
  {Durant}}, \bibinfo {author} {\bibfnamefont {J.}~\bibnamefont {Morokuma}},
  \bibinfo {author} {\bibfnamefont {C.}~\bibnamefont {Fields}}, \bibinfo
  {author} {\bibfnamefont {K.}~\bibnamefont {Williams}}, \bibinfo {author}
  {\bibfnamefont {D.~S.}\ \bibnamefont {Adams}}, \ and\ \bibinfo {author}
  {\bibfnamefont {M.}~\bibnamefont {Levin}},\ }\href {\doibase
  https://doi.org/10.1016/j.bpj.2017.04.011} {\bibfield  {journal} {\bibinfo
  {journal} {Biophysical Journal}\ }\textbf {\bibinfo {volume} {112}},\
  \bibinfo {pages} {2231} (\bibinfo {year} {2017})}\BibitemShut {NoStop}%
\bibitem [{\citenamefont {Yamamoto}\ \emph {et~al.}(2019)\citenamefont
  {Yamamoto}, \citenamefont {Bone}, \citenamefont {Sohn}, \citenamefont {Syed},
  \citenamefont {Reissaus}, \citenamefont {Mosley}, \citenamefont {Wijeratne},
  \citenamefont {True}, \citenamefont {Tong}, \citenamefont {Kono},\ and\
  \citenamefont {Evans-Molina}}]{Yamamoto_JCB_2021}%
  \BibitemOpen
  \bibfield  {author} {\bibinfo {author} {\bibfnamefont {W.~R.}\ \bibnamefont
  {Yamamoto}}, \bibinfo {author} {\bibfnamefont {R.~N.}\ \bibnamefont {Bone}},
  \bibinfo {author} {\bibfnamefont {P.}~\bibnamefont {Sohn}}, \bibinfo {author}
  {\bibfnamefont {F.}~\bibnamefont {Syed}}, \bibinfo {author} {\bibfnamefont
  {C.~A.}\ \bibnamefont {Reissaus}}, \bibinfo {author} {\bibfnamefont {A.~L.}\
  \bibnamefont {Mosley}}, \bibinfo {author} {\bibfnamefont {A.~B.}\
  \bibnamefont {Wijeratne}}, \bibinfo {author} {\bibfnamefont {J.~D.}\
  \bibnamefont {True}}, \bibinfo {author} {\bibfnamefont {X.}~\bibnamefont
  {Tong}}, \bibinfo {author} {\bibfnamefont {T.}~\bibnamefont {Kono}}, \ and\
  \bibinfo {author} {\bibfnamefont {C.}~\bibnamefont {Evans-Molina}},\ }\href
  {\doibase 10.1074/jbc.ra118.005683} {\bibfield  {journal} {\bibinfo
  {journal} {Journal of Biological Chemistry}\ }\textbf {\bibinfo {volume}
  {294}},\ \bibinfo {pages} {168} (\bibinfo {year} {2019})}\BibitemShut
  {NoStop}%
\end{thebibliography}%
\bibliographystyle{apsrev4-1}



\end{document}